\documentclass[pra,onecolumn,showpacs,preprintnumbers,amsmath,amssymb]{revtex4}
\usepackage{graphicx}
\usepackage{dcolumn}
\usepackage{bm}

\begin{document}

\title{Diffraction of light by a planar aperture in a metallic screen}

\author{A.~Drezet }
\email{aurelien.drezet@uni-graz.at} \affiliation{Institute of
Physics, Karl Franzens Universit\"at Graz, Universit\"atsplatz 5
A-8010 Graz, Austria}
\author{J.~C.~Woehl  and S.~Huant }
\affiliation{Laboratoire de Spectrom\'{e}trie Physique, CNRS
UMR5588, Universit\'{e} Joseph Fourier Grenoble, BP 87, 38402
Saint Martin d'H\`{e}res cedex, France}
\date{received 25 February 2006 accepted 24 march 2006 }

\begin{abstract}
We present a complete derivation of the formula of Smythe
[Phys.~Rev.~72, 1066 (1947)] giving the electromagnetic field
diffracted by an aperture created in a perfectly conducting plane
surface. The reasoning, valid for any excitating field and any
hole shape, makes use only of the free scalar Green function for
the Helmoltz equation without any reference to a Green dyadic
formalism. We compare our proof with the one previously given by
Jackson and connect our reasoning to the general Huygens Fresnel
theorem.
\end{abstract}

\maketitle

\section{Introduction}
 Diffraction of electromagnetic waves by an aperture in a perfect
metallic plane is not only a mathematical problem
 of fundamental interest but is connected to many applications in the microwave domain (for example, in waveguides
 and in cavity resonators~\cite{Collin}) as well as in the optical
regime where it is involved in many optical
arrangements~\cite{Born}. The fundamental importance of this
phenomenon in near-field optics has been pointed out as early as
in 1928 by Synge~\cite{Synge} in his prophetic paper and is
currently involved in modern near-field scanning
optical microscopy (NSOM)~\cite{Pohl}.\\
 In the domain of
applicability of NSOM where distances and dimensions are smaller
than or close to the wavelength of light, we need  to know the
exact structure of the electromagnetic field, and we cannot in
general consider the usual approximations involved in Kirchhoff's
theory for a scalar wave~\cite{Drezet1,Drezet2,Drezet3}. In this
context, one of the most cited approaches is the one given by
Bethe~\cite{Bethe} in 1944 and corrected by
Bouwkamp~\cite{Bouwkamp1,Bouwkamp2}. It gives the electromagnetic
field diffracted by a small circular aperture in a perfect
metallic plane in the limit where the optical wavelength is much
larger than the aperture. Less known is the more general formula
of Smythe~\cite{Smythe,Smythe2} which expresses in a formal way
the Huygens Fresnel principle for any kind of aperture in a
metallic screen. Even if this formula is not an explicit solution
for the general diffraction problem, it constitutes an integral
equation which can be used in a self consistent way in
perturbative or numerical calculations of the diffracted
field~\cite{Butler,Eggimann}. Further efforts have been made by
Smythe~\cite{Smythe,Smythe2} himself in order to justify his
formula by means of some arrangements of current sheets fitting
the aperture. This method essentially consists of transforming the
problem of diffraction by a hole into a physically different one
in order to guess the correct integral equation for the original
problem. However, if this physical reasoning proves the
consistency of the proposed solution with Maxwell equations and
boundary conditions for the field, it is not directly connected to
the rigorous electromagnetic
 formulation of the Huygens Fresnel principle obtained by Stratton and Chu ~\cite{Stratton}. Such a connection
 is expected naturally because these two formulations of
diffraction must be equivalent here.\\
Jackson ~\cite{Jackson}, in the first edition
 of his textbook on electrodynamics, developed a complete
 proof of the Smythe formula starting from the Stratton and Chu formula [Eq.~(\ref{supergreen}) of the present
 paper]. Nevertheless, like in the original paper of Smythe,
 Jackson transforms the problem into a physically different one in order to guess the correct result.
  The result is then subjected to the same remarks as above for Smythe's approach. Other justifications of
Smythe results are based on the use of the Babinet theorem or of
the Green dyadic method. The latter, which uses a tensorial Green
function instead of a scalar one like in Kirchhoff's or Stratton
and Chu's theories, gives us the most direct justification for
Smythe approach in terms of the Huygens Fresnel principle.
However, this proof is for the moment not directly
connected to the Stratton and Chu approach. It is the aim of this paper to establish such a link.\\
The paper is organized as follows. We give in Sec. II a
description of the general theory of diffraction of
electromagnetic waves by an aperture in a screen. In Sec. III, we
exploit precedent works by Jackson ~\cite{Jackson,Jackson2} and
Levine and Schwinger ~\cite{Levine} to justify directly and
rigorously the Smythe formula using the Stratton Chu theorem
without relying on any ingenious physical ``trick''. Sec. IV deals
with a vectorial justification of Smythe's approach. The
consistency between the various theoretical treatments of
diffraction by an aperture in a metallic screen is stressed in
Sec. V which also compares our treatment with that obtained within
the Green dyadic formalism ~\cite{Schwinger1,Schwinger2}. Our
conclusions appear in Section VI.

\section{The diffraction problem in electromagnetism}
The first coherent theory of diffraction was elaborated by
Kirchhoff (1882) on the basis of the
 Huygens Fresnel principle ~\cite{Born,Poincare}. The method of integral
 equations allows one to write a solution $\psi\left(\vec{r}\right)e^{-i\omega
 t}$ of the Helmholtz propagation equation $[\nabla^{2}+k^{2}]
 \psi\left(\vec{r}\right)=0$ ($k=\omega/c$) using the  ``free" scalar Green function
$G \left(\vec{r},\vec{r'}\right)= e^{ikR}/4\pi
 R$ which is a solution of the equation $[\nabla^{2}+k^{2}]
G\left(\vec{r},\vec{r'}\right)=-\delta^{3}\left(\vec{r}-\vec{r'}\right)$.\\
If, as schematized in Fig.~1, we consider now an aperture  $\delta
S$ made in a two-dimensional infinite screen $S$ and illuminated
by incident radiation, we can express the field $\psi$ existing at
each observation point located behind the screen (i.~e.~, for
$z>0$) by the Kirchhoff formula
\begin{equation}
\psi\left(\vec{r}\right) =\int_{S}[
\psi\left(\vec{r'}\right)\vec{n'}\cdot\overrightarrow{\nabla'}G\left(\vec{r},\vec{r'}\right)
-G\left(\vec{r},\vec{r'}\right)\vec{n'}\cdot\overrightarrow{\nabla'}\psi\left(\vec{r'}\right)]\textrm{d}S',
\label{green}
\end{equation}
where the normal unit vector ${n'}$ is oriented into the
diffraction half-space.\\
In a problem of diffraction, we usually impose the additional
first Kirchhoff  ``shadow" approximation
$\psi\left(\vec{r'}\right)=\partial_{n'}\psi\left(\vec{r'}\right)=0$
which is valid on the unilluminated side of the screen. This
permits one to restrict the integral in (\ref{green}) to the
region of the aperture only, which is very useful in some
approximations or iterative resolutions. Nevertheless, this
intuitive hypothesis has some fundamental inconsistencies because,
following a theorem due to Poincar\'{e}~\cite{Poincare}, a field
satisfying the shadow approximation on a finite domain must vanish
everywhere.\\
A classic solution proposed by Rayleigh~\cite{Rayleigh} and
Sommerfeld~\cite{Sommerfeld} to circumvent this difficulty
consists in
 replacing the free Green function by the Dirichlet
$G_{D}$ or the Neumann $G_{N}$ Green functions~\cite{Jackson}
satisfying $\partial_{n'}G_{N} \left(\vec{r},\vec{r'}\right)=0$
and $G_{D} \left(\vec{r},\vec{r'}\right)=0$ for all points
$\vec{r'}$ on $S$. We can then rigourously reduce the integral to
the region of the aperture depending on the nature of the boundary
problem. For example, if we impose $\psi=0$ on the screen, we can
then write
\begin{equation}
\psi\left(\vec{r}\right) =\int_{\textrm{Aperture}}
\psi\left(\vec{r'}\right)
 \partial_{n'}G_{D}\left(\vec{r},\vec{r'}\right)\textrm{d}S'.\label{rayleigh}
 \end{equation}
In principle, it could be possible to generalize the preceding
methods to the different Cartesian components $\psi_{\alpha}$ of
the electromagnetic field using equations of the form
$\psi_{\alpha}=\int_{S}[\psi_{\alpha}\partial_{n'}G-G\partial_{n'}\psi_{\alpha}]\textrm{d}S'$.
Nevertheless, as pointed out by Stratton, Chu ~\cite{Stratton} and
others ~\cite{Love,Larmor,Kottler}, the Maxwell equations couple
the field components between them and the consistency of these
relations must be controlled if we use an integral equation like
Eq.~(\ref{green}) either in an exact or approximative treatment of
diffraction. In addition, because the boundary conditions imposed
by Maxwell's equations connect the tangential and the normal
components of the field  on the  screen surface, it is not at all
trivial to reduce the integral to the region of the aperture
directly using Eq.~(\ref{green}).\\ Due to the uniqueness theorem,
such possible reduction of the integral appearing in the Huygens
Fresnel principle is expected in the case of a perfectly
conducting metallic screen. Indeed, following this uniqueness
theorem, the field in the diffracted space must depend only on the
tangential electric field on the screen and aperture surface.
Because the tangential electric field vanishes on the screen, the
integral must depend only on the tangential field
at the opening.\\
 Numerous authors, especially Stratton and
Chu ~\cite{Stratton} as well as Schelkunoff
~\cite{Schelkunoff,Schelkunoff2}, have discussed a vectorial
integral equation satisfying Maxwell's equations automatically. We
can effectively write
\begin{eqnarray}
\overrightarrow{E}\left(\vec{x}\right)=\int_{S}[ik\left(\vec{n'}\times
\overrightarrow{B}\right)G+
\left(\vec{n'}\times\overrightarrow{E}\right)\times\overrightarrow{\nabla'}G+\nonumber\\\left(\vec{n'}\cdot\overrightarrow{E}\right)\overrightarrow{\nabla'}G]\textrm{d}S',
\label{supergreen}
\end{eqnarray}
hereafter referred to as the Stratton Chu equation. A similar
expression holds for the magnetic field by means of the
substitution $\overrightarrow{E}\rightarrow\overrightarrow{B}$ and
$\overrightarrow{B}\rightarrow
 -\overrightarrow{E}$.\\
 It is important to note that
Eq.~(\ref{supergreen}) is over-determined although it depends
explicitly on the tangential and normal components of the
electromagnetic field defined on $S$. Indeed, due to the
equivalence principle of Love and
 Schelkunoff ~\cite{Love,Schelkunoff,Stratton2} and to the uniqueness theorem, we expect that the ``most adapted"
integral equations depend only on
$\vec{n'}\times\overrightarrow{E}$
  \emph{or} $\vec{n'}\times\overrightarrow{B}$ on $S$. In addition, unlike in the scalar case, we cannot
 directly reduce the surface integral to the region of the aperture
 just by choosing an adapted Dirichlet or Neumann Green function.
 It seems then necessary to apply once again the shadow approximation of
 Kirchhoff in order to simplify the integration despite the
 inconsistency of the method. As in the Poincar\'e theorem, some problems appear here because we need to add
a nonphysical contour integral associated with a magnetic line
charge in Eq.~(\ref{supergreen}) (or to an electric line charge in
the equivalent formula for ${B}$) in order to satisfy Maxwell's
equations and to compensate for the arbitrary change imposed to
the integration domain~\cite{Meixner}.
 Furthermore, in this Kirchhoff Kottler~\cite{Kottler} theory, the introduction
of contour integrals
induces a logarithmic divergence of the energy at the rim of the aperture, a fact which is forbidden in a diffraction problem.\\
The particular case of the diffraction by an aperture in a planar
screen constitutes an exception in the sense that a rigorous
integral equation had been anticipated  by Schelkunoff
~\cite{Schelkunoff} and Bethe ~\cite{Bethe} for a subwavelength
circular aperture
 and
generalized by Smythe ~\cite{Smythe,Smythe2} for any kind of
aperture. The integral equation is
\begin{eqnarray}
 \overrightarrow{E}\left(\vec{x}\right)
 =\frac{1}{2\pi}\overrightarrow{\nabla}\times\Big(\int_{\texttt{Aperture}}
 \left({\hat{z}}\times\overrightarrow{E}\right)
 \frac{e^{ikR}}{R}\textrm{d}S'\Big). \label{smythe0}
\end{eqnarray}
For some applications, it is important to note that in the short
wavelength limit ($\lambda\ll \textrm{aperture typical radius}$)
for which the electromagnetic field in the aperture can be
identified with the incident plane wave
$\overrightarrow{B}_{i}={\hat{z}}\times\overrightarrow{E}_{i}$
(first Kirchhoff approximation), the formula of Stratton Chu
limited to the aperture domain and the exact solution of Smythe
give approximately the same result. Indeed, within the Fraunhofer
approximation, Eq.~(\ref{smythe0}) reads
\begin{equation}
\overrightarrow{E}\simeq\frac{ike^{ikr}}{
r}{\hat{r}}\times\int_{\texttt{Aperture}}\Big(\frac{{\hat{z}}\times\overrightarrow{E}_{i}}{2\pi}
e^{-ik{\hat{r}}\cdot\overrightarrow{x'}}\Big)\textrm{d}S',\label{aleph}
\end{equation}  whereas Eq.~(\ref{supergreen}) reduces to
\begin{equation}
\overrightarrow{E}\simeq\frac{ike^{ikr}}{
r}\frac{{\hat{r}}+{\hat{z}}
}{2}\times\int_{\texttt{Aperture}}\Big(\frac{{\hat{z}}\times
\overrightarrow{E}_{i}}{2\pi}
e^{-ik{\hat{r}}\cdot\vec{x'}}\Big)\textrm{d}S'.
\end{equation} Both equations are identical in the practical limit of small
diffraction angles, i.~e.~, close to the normal axis $z$ going
through the aperture. Equation (\ref{aleph}) is correct for a
subwalength aperture only because we cannot identify the field in
the aperture with the incident one. We can see that the asymptotic
diffracted field for $z>0$ is equivalent to the one produced by an
effective magnetic dipole
\begin{equation}
\overrightarrow{M}_{\textrm{eff}}=\int_{\texttt{Aperture}}\Big(\frac{\vec{n'}\times\overrightarrow{E}}{2\pi
ik} \Big)\textrm{d}S',\label{dipôlemag}
 \end{equation}
 and by an effective electric dipole
\begin{equation}
\overrightarrow{P}_{\textrm{eff}}
=\frac{{\hat{z}}}{4\pi}\int_{\texttt{Aperture}}\left(\vec{x'}\cdot\overrightarrow{E}\right)\textrm{d}S'.\label{dipôleelec}
\end{equation} These formula are fundamental in the context of
NSOM because they give us the Bethe
Bouwkamp~\cite{Bethe,Bouwkamp1,Bouwkamp2,Jackson} dipoles which,
in the particular case of a circular aperture of radius $a$, are
\begin{equation}
\overrightarrow{P}_{\textrm{eff}}=\frac{a^{3}}{3\pi}\overrightarrow{E}_{\bot}^{\left(0\right)}
\quad,
\overrightarrow{M}_{\textrm{eff}}=-\frac{2a^{3}}{3\pi}\overrightarrow{B}_{\|}^{\left(0\right)}.
\label{rayleighdipôle}
\end{equation}
$\overrightarrow{E}_{\bot}^{\left(0\right)}$ and
$\overrightarrow{B}_{\|}^{\left(0\right)}$ are, respectively, the
locally uniform normal electric field and tangential magnetic
field existing in the aperture zone in the absence of the opening
(in $z=0^{-}$).

\section{Green dyadic justification of the Smythe formula}

 The so-called Smythe formula Eq.~(\ref{smythe0}) is generally obtained on the basis
  of different principles such as the Babinet principle or the equivalence theorem (see Schelkunoff ~\cite{Schelkunoff},
Bouwkamp ~\cite{Bouwkamp3}, Jackson ~\cite{Jackson2}). In
particular, the equivalence theorem shows that the solution of
Smythe for $z>0$ is identical to the one obtained by considering a
virtual surface magnetic-current density given by
$\overrightarrow{J}_{s}^{m}=-c{\hat{z}}\times\overrightarrow{E}/\left(2\pi\right)$.
All these derivations are self consistent if we consider the very
fact that the guessed results fulfill Maxwell equations. Then, the
uniqueness theorem ensures that the result is the only one
possible. Nevertheless, as already noted, the calculation is not
direct and not necessarily connected to the Stratton and Chu
formalism. A classical calculation due to Schwinger and Levine
~\cite{Schwinger1,Schwinger2} shows, however, that it is possible
to rigourously and directly obtain this equation using the
tensorial, or dyadic,
 Green function formalism.\\
  Such an electric dyadic Green function ~\cite{Chentotai}
  $\stackrel{\leftrightarrow}{G}_{e}$, which is solution of the equation
  \begin{eqnarray}
\overrightarrow{\nabla}\times\left(\overrightarrow{\nabla}\times
\stackrel{\leftrightarrow}{G}_{e}
\left({r},{r'}\right)\right)=k^{2}\stackrel{\leftrightarrow}{G}_{e}\left(\vec{r},\vec{r'}\right)+
\stackrel{\leftrightarrow}{\delta} \delta^{3} \left(\vec{r}-
\vec{r'}\right)
\end{eqnarray} (with $\stackrel{\leftrightarrow}{\delta}=\sum_{i}  {\hat{x}}_{i} {\hat{x}}_{i} $)
 satisfying the condition
 $\overrightarrow{\nabla}\cdot\stackrel{\leftrightarrow}{{G}}_{e}=-\left(1/k^{2}\right)\overrightarrow{\nabla}
\delta^{3} \left(\vec{r}-\vec{r'}\right) $, can be used to write
the integral equation
 \begin{equation}
\overrightarrow{E}\left(\vec{r}\right)=\int_{S}[\left(\vec{n'}\times\overrightarrow{E}\right)
\cdot\overrightarrow{\nabla'}\times\stackrel{\leftrightarrow}{{G}}_{e}-ik\overrightarrow{B}\cdot\left(\vec{n'}\times
\stackrel{\leftrightarrow}{{G}}_{e}\right)  ]\textrm{d}S'
\end{equation}
which is defined on the same surface as previously. By imposing
the dyadic Dirichlet condition $\vec{n'}\times
\stackrel{\leftrightarrow}{{G}}_{e}=0$ on $S$, we can obtain the
relation
\begin{equation}
\overrightarrow{E}\left(\vec{r}\right)=\int_{\texttt{Aperture}}[\left(\vec{n'}\times\overrightarrow{E}\right)\cdot\overrightarrow{\nabla'}\times\stackrel{\leftrightarrow}{{G}}_{e}
]\textrm{d}S'\label{dyadicdic}
\end{equation} which depends only on the tangential electric field
at the aperture. This is in perfect agreement with the equivalence
principle and the uniqueness theorem.\\Following
Ref.~~\cite{Chentotai}, the total
 Green function $\stackrel{\leftrightarrow}{{G}}_{e}$ for the plane can be deduced from
 the ``free" dyadic \begin{eqnarray}
\stackrel{\leftrightarrow}{{G}}_{e}^{0}\left(\vec{r},\vec{r'}\right)=\left(\stackrel{\leftrightarrow}{\delta}
+\frac{1}{k^{2}}\overrightarrow{\nabla}\overrightarrow{\nabla}\right)\frac{e^{ikR}}{4\pi
R}
\end{eqnarray} [with $R=\sqrt{ \left(x-x'\right)^{2} + \left(y-y'\right)^{2}+
\left(z-z'\right)^{2}}$] by using the image method. We have
\begin{eqnarray}
\stackrel{\leftrightarrow}{{G}}_{e}
\left(\vec{r},\vec{r'}\right)=\left(\stackrel{\leftrightarrow}{\delta}-\frac{1}{k^{2}}\overrightarrow{\nabla}
\overrightarrow{\nabla'}\right) G_{D}\left(\vec{r},\vec{r'}\right)
+2{\hat{z}}{\hat{z}}\frac{e^{ikR'}}{4\pi R'},\label{dyade}
 \end{eqnarray}
where $G_{D}=\left(e^{ikR}/R-e^{ikR'}/R'\right)/4\pi$  is the
scalar Dirichlet Green function for the plane screen, and
 $R'=\sqrt{ \left(x-x'\right)^{2} + \left(y-y'\right)^{2}+
\left(z+z'\right)^{2}}$. Inserting this Green function into
Eq.~(\ref{dyadicdic}) gives us directly Eq.~(\ref{smythe0}). It is
interesting to observe that with the Green dyadic method,  we can
recover the formula of Smythe by using a magnetic current
distribution located in front of a metallic plane
or, equivalently, by using a double layer of magnetic currents propagating in the same direction~\cite{Butler}. \\
 In theory, both approaches based either on the scalar Green functions or on the dyadic Green functions are
equivalent. In practice however, the
 difficulties related to the Stratton Chu formula
 Eq.~(\ref{supergreen}) have imposed the Green dyadic
 method. An illustration of this statement is that the dyadic formalism has been
 extensively used in the context of the electromagnetic theory of
NSOM~\cite{Girard1,Girard2,Girard3,Bozhe}.
\section{Vectorial justification of the Smythe formula}

 We propose now a justification of Eq.~(\ref{smythe0}) based on the
Stratton Chu formula Eq.~(\ref{supergreen}). This derivation will
directly reveal the equivalence of the scalar and dyadic
approaches
in the particular case of a planar screen with an aperture.\\
 Let the surface $S$ of equation $z=0$ be an infinite, perfectly conducting
 metallic screen containing an aperture covering the surface $\delta
 S$.  By the definition of diffraction, we can always separate
 the total electric (magnetic) field $\overrightarrow{E}$ ($\overrightarrow{B}$) into an incident field $\overrightarrow{E}^{i}$
 ($\overrightarrow{B}^{i}$ ) existing independently of the presence of the screen, and into a
 diffracted field $\overrightarrow{E}'$ ($\overrightarrow{B}'$) produced by  the surface charge and current densities
 $\rho'_{s}, \overrightarrow{J'}_{s}$ located on the metal.\\
 We have
 $\overrightarrow{B'}=\overrightarrow{\nabla}\times\overrightarrow{A'}$ and
$\overrightarrow{E'}=-\overrightarrow{\nabla}\Phi'+ik\overrightarrow{A'}
$ where potentials are expressed in a Lorentz gauge
\begin{eqnarray}
\overrightarrow{A'}\left(\vec{r}\right)=\int_{\textrm{Screen}}\textrm{d}S'
\Big(\frac{\overrightarrow{J'}_{s}}{c}\left(\vec{r'}\right)\frac{e^{iKR}}{R}\Big),\nonumber \\
\Phi'\left(\vec{r}\right)
=\int_{\textrm{Screen}}\textrm{d}S'\Big(\rho'_{s}\left(\vec{r'}\right)\frac{e^{iKR}}{R}\Big),\nonumber\\
\end{eqnarray}
with  $R=\|\vec{r}-\vec{r'}\|$ (we omit here the time dependent
factor $e^{-i\omega t}$). Because these potentials are even
functions of $z$ we then have the  following symmetries
\begin{eqnarray}
\textrm{$E'_{x}$, $E'_{y}$, $B'_{z}$ are even in $z$,}\nonumber\\
\textrm{$E'_{z}$, $B'_{x}$, $B'_{y}$ are odd in $z$. }
\label{symetrie}
\end{eqnarray}
These symmetries already used by Jackson ~\cite{Jackson,Jackson2}
imply in particular $E'_{z}=B'_{y}=B'_{x}=0$ at the aperture.
Therefore, the field is a discontinuous function through
the metal.\\
 Let us now consider an observation point ${x}$
located in the half space $z>0$.
 We can apply the vectorial Green theorem on a closed integration surface made up of a half sphere
  $ S_{\infty}^{+}$  ``at infinity" and of
the $S^{+}$ plane ($z=0^{+}$) as seen in Fig.~2 (A). This surface
$S^{+}$ can itself be decomposed into an aperture region $\delta
S^{+}$ and
into a screen region $\left(S-\delta S\right)^{+} $.\\
We have then
\begin{widetext}
\begin{eqnarray}
\overrightarrow{E'}\left(\vec{x}\right)=\int_{\left(S-\delta
S\right)^{+}}[ik\left(\vec{n'}\times\overrightarrow{B'}\right)G
+\left(\vec{n'}\times\overrightarrow{E'}\right)\times\overrightarrow{\nabla'}G
+
\left(\vec{n'}\cdot\overrightarrow{E'}\right)\overrightarrow{\nabla'}G]\textrm{d}S'
 +\int_{\delta
S^{+}}[\left(\vec{n'}\times\overrightarrow{E'}\right)\times\overrightarrow{\nabla'}G]\textrm{d}S'\nonumber\\
+\int_{S_{\infty}^{+}}[ik\left(\vec{n'}\times\overrightarrow{B'}\right)G+\left(\vec{n'}\times\overrightarrow{E'}\right)\times\overrightarrow{\nabla'}G
+\left(\vec{n'}\cdot\overrightarrow{E'}\right)\overrightarrow{\nabla'}G]\textrm{d}S',\nonumber\\
 \label{smythe1}
\end{eqnarray}
\end{widetext}
where the unit vector $\vec{n'}$ lies on $S^{+}$  and is oriented
in the positive z direction: $\vec{n'}={\hat{z}}$. Similarly we
can consider the surface of integration represented in Fig.~2 (B).
We obtain an integration on the $S^{+}_{\infty}$, $S^{-}_{\infty}$
surfaces
 and on  $\left(S-\delta S\right)^{+} $  and $\left(S-\delta S\right)^{-} $ surfaces. Such integration surfaces have
already been used by Schwinger
 and Levine in the context of diffraction by a scalar
wave ~\cite{Levine}. Here, due to the symmetries
  given by Eq.~(\ref{symetrie}), we deduce
\begin{widetext}
\begin{eqnarray}
\overrightarrow{E'}\left(\vec{x}\right)=2\int_{\left(S-\delta
S\right)^{+}}[ik\left(\vec{n'}\times\overrightarrow{B'}\right)G+
\left(\vec{n'}\cdot\overrightarrow{E'}\right)\overrightarrow{\nabla'}G]\textrm{d}S'
+\int_{S_{\infty}^{-}}[ik\left(\vec{n'}\times\overrightarrow{B'}\right)G+\left(\vec{n'}\times\overrightarrow{E'}\right)\times\overrightarrow{\nabla'}G+
\left(\vec{n'}\cdot\overrightarrow{E'}\right)\overrightarrow{\nabla'}G]\textrm{d}S'.
\nonumber\\+\int_{S_{\infty}^{+}}[ik\left(\vec{n'}\times\overrightarrow{B'}\right)G+\left(\vec{n'}\times\overrightarrow{E'}\right)\times\overrightarrow{\nabla'}G+
\left(\vec{n'}\cdot\overrightarrow{E'}\right)\overrightarrow{\nabla'}G]\textrm{d}S'\nonumber\\
\label{smythe2}
\end{eqnarray}
\end{widetext}
with $\vec{n'}={\hat{z}}$ on the $\left(S-\delta S\right)^{+}$
surface. After identification of Eq.~(\ref{smythe1}) and
Eq.~(\ref{smythe2}), we obtain
\begin{widetext}
\begin{eqnarray}
\overrightarrow{E'}\left(\vec{x}\right)=2\int_{
S^{+}}[\left(\vec{n'}\times\overrightarrow{E'}\right)\times\overrightarrow{\nabla'}G]\textrm{d}S'
-\int_{S_{\infty}^{-}}[ik\left(\vec{n'}\times\overrightarrow{B'}\right)G+\left(\vec{n'}\times\overrightarrow{E'}\right)\times\overrightarrow{\nabla'}G+
\left(\vec{n'}\cdot\overrightarrow{E'}\right)\overrightarrow{\nabla'}G]\textrm{d}S'
\nonumber\\+\int_{S_{\infty}^{+}}[ik\left(\vec{n'}\times\overrightarrow{B'}\right)G+\left(\vec{n'}\times\overrightarrow{E'}\right)\times\overrightarrow{\nabla'}G+
\left(\vec{n'}\cdot\overrightarrow{E'}\right)\overrightarrow{\nabla'}G]\textrm{d}S'.\nonumber\\
\label{smythe3}
\end{eqnarray}
\end{widetext}
In order to simplify this formula, it is important to note that
the fields $\overrightarrow{E'}$, $\overrightarrow{B'}$ located on
$S_{\infty}^{\pm}$ are the reflected fields
$\overrightarrow{E}^{r}$, $\overrightarrow{B}^{r}$ which could be
produced by the complete metallic screen $z=0$  submitted to the
same incident field in the absence of
the aperture. \\
 Because this field compensates for the incident field for $z>0$,
 we have  $\overrightarrow{E}^{r}=-\overrightarrow{E}^{i}$,
$ \overrightarrow{B}^{r}=-\overrightarrow{B}^{i}$ in this
half-space. As a consequence, the integral on $S_{\infty}^{+}$ in
Eq.~(\ref{smythe3}) can be written\\
  $ -\overrightarrow{E}^{i}\left(\vec{x}\right)+
 \int_{S^{+}}[ik\left(\vec{n'}\times\overrightarrow{B}^{i}\right)G+\left(\vec{n'}\times\overrightarrow{E}^{i}\right)\times\overrightarrow{\nabla'}G+
\left(\vec{n'}\cdot\overrightarrow{E}^{i}\right)\overrightarrow{\nabla'}G]\textrm{d}S'$,
which is a direct application  of the Green theorem for an
observation point located on the closed surface composed of
$S_{\infty}^{+}$ and $S^{+}$.\\
 Injecting this last result into
Eq.~(\ref{smythe3}) and after subtracting and adding $2\int_{
S^{+}}[\left(\vec{n'}\times\overrightarrow{E}^{i}\right)\times\overrightarrow{\nabla'}G]\textrm{d}S'$,
we finally obtain
$\overrightarrow{E'}=\overrightarrow{E}^{\left(1\right)}+\overrightarrow{E}^{\left(2\right)}$
where
\begin{eqnarray}
\overrightarrow{E}^{\left(1\right)}\left(\vec{x}\right)=2\int_{
S^{+}}[\left(\vec{n'}\times\overrightarrow{E}\right)\times\overrightarrow{\nabla'}G]\textrm{d}S'-\overrightarrow{E}^{i}\left(\vec{x}\right)\textrm{d}S'\end{eqnarray}
and
\begin{widetext}
\begin{eqnarray}
\overrightarrow{E}^{\left(2\right)}\left(\vec{x}\right)=-\int_{S_{\infty}^{-}}[ik\left(\vec{n'}\times\overrightarrow{B}^{r}\right)G
+\left(\vec{n'}\times\overrightarrow{E}^{r}\right)\times\overrightarrow{\nabla'}G+
\left(\vec{n'}\cdot\overrightarrow{E}^{r}\right)\overrightarrow{\nabla'}G]\textrm{d}S'.
\nonumber\\+\int_{S^{+}}[ik\left(\vec{n'}\times\overrightarrow{B}^{i}\right)G-\left(\vec{n'}\times\overrightarrow{E}^{i}\right)\times\overrightarrow{\nabla'}G+
\left(\vec{n'}\cdot\overrightarrow{E}^{i}\right)\overrightarrow{\nabla'}G]\textrm{d}S'.\nonumber\\
\label{smythe4}
\end{eqnarray}
\end{widetext}
Because of Eq.~(\ref{symetrie}), we also have
\begin{eqnarray}
E^{r}_{x,y}\left(x,y,z\right)= -E^{i}_{x,y}\left(x,y,-z\right),\nonumber\\
  B^{r}_{ z}\left(x,y,z\right)= -B^{i}_{
  z}\left(x,y,-z\right),\nonumber
  \end{eqnarray} and \begin{eqnarray}
B^{r}_{ x,y}\left(x,y,z\right)= B^{i}_{ x,y}\left(x,y,-z\right),\nonumber\\
 E^{r}_{ z}\left(x,y,z\right)= E^{i}_{ z}\left(x,y,-z\right)\nonumber\\
 \label{symetrie2}
\end{eqnarray}
for $z<0$.
 Using the fact that the integral on $S^{+}$ can be written as an integral on $S^{-}$:
$\int_{S^{+}}\{
\overrightarrow{E}^{i},\overrightarrow{B}^{i}\}=-\int_{S^{-}}\{
\overrightarrow{E}^{i},\overrightarrow{B}^{i}\}$, and using
Eq.~(\ref{symetrie2}) , the last two integrals in
Eq.~(\ref{smythe4}) can be transformed into
$\int_{S^{-}+S^{-}_{\infty}}[ik\left(\vec{n'}\times\overrightarrow{B}^{r}\right)G+\left(\vec{n'}\times\overrightarrow{E}^{r}\right)\times\overrightarrow{\nabla'}G+
\left(\vec{n'}\cdot\overrightarrow{E}^{r}\right)\overrightarrow{\nabla'}G]\textrm{d}S'$.
Because the observation point is outside of the closed surface
composed of $S_{\infty}^{-}$ and of $S^{-}$,
$\overrightarrow{E}^{\left(2\right)}\left(\overrightarrow{x}\right)$
is
 zero.
 Regrouping all terms, the total electric field in the half plane $z>0$ is finally given
 by the Smythe formula:
\begin{eqnarray}
 \overrightarrow{E}\left(\overrightarrow{x}\right)= 2\int_{\delta S^{+}}[\left(\vec{n'}\times\overrightarrow{E}\right)
 \times\overrightarrow{\nabla'}G]\textrm{d}S' \nonumber\\ =\frac{1}{2\pi}\overrightarrow{\nabla}\times\Big(\int_{\textrm{Aperture}}\left({\hat{z}}\times\overrightarrow{E}\right)
 \frac{e^{ikR}}{R}\textrm{d}S'\Big) \label{smythe5}
\end{eqnarray}where we have applied Maxwell's boundary conditions that
annihilate  the tangential component of the total electric field
on a perfect metal.
 An equivalent derivation in the $z<0$ half
space gives
\begin{eqnarray}
 \overrightarrow{E}\left(\overrightarrow{x}\right)= \overrightarrow{E}^{0}\left(\overrightarrow{x}\right)+2\int_{\delta S^{-}}[\left(\vec{n'}\times\overrightarrow{E}\right)
 \times\overrightarrow{\nabla'}G]\textrm{d}S' \nonumber\\ = \overrightarrow{E}^{0}\left(\overrightarrow{x}\right)-\frac{1}{2\pi}\overrightarrow{\nabla}\times\Big(\int_{\textrm{Aperture}}
  \left({\hat{z}}\times\overrightarrow{E}\right)\frac{e^{ikR}}{R}\textrm{d}S'\Big), \label{smythe6}
\end{eqnarray}
where $\overrightarrow{E}^{0}=
\overrightarrow{E}^{i}+\overrightarrow{E}^{r}$ is now the total
electric field existing in the $z<0$ domain for the problem
without aperture.

\section{Consistency between various approaches}

As written in the introduction, the proof given by Jackson
~\cite{Jackson} of the Smythe equation is connected to the theory
of vectorial diffraction Eq.~(\ref{supergreen}). In order to solve
the problem, Jackson used a volume looking like a flat pancake
limited by the two $S^{+}$ and $S^{-}$ surfaces, and he applied
Eq.~(\ref{supergreen}) to this boundary. Then, in agreement with
Smythe, Jackson imagined a double current sheet such that the
surface current on the two $S^{+}$ and $S^{-}$ layers at any point
of a given area fitting the aperture are equal and opposite. With
such a distribution, it is possible to reduce the integral of
Eq.~(\ref{supergreen}) to the one given by the formula of Smythe,
Eq.~(\ref{smythe5}). Such a formula is then the correct one to
describe the diffraction problem by an aperture in agreement with
the uniqueness theorem.\\
Our justification of the Smythe theorem is more direct because it
uses only the Huygens Fresnel theorem without applying the
intuitive trick of a virtual surface current distribution
associated with a different physical situation (double layer of
electric current, or layer of magnetic current confined to the
aperture zone). Our result is in fact the direct generalization of
a method used by the authors for a scalar wave $\psi$. Using two
different surface integrations, as the ones used in this paper, we
are indeed able to prove directly the Rayleigh-Sommerfeld theorem
given by Eq.~(\ref{rayleigh}). This scalar reasoning, which is
similar to the one presented before, is given in the appendix. It
can be observed that the scalar result makes only use of the Green
function in vacuum $G$ in order to justify the result obtained
with the Dirichlet one $G_{D}$. Similarly, our derivation of the
Smythe formula uses the scalar Green function in order to justify
the result obtained with the ``Dirichlet'' dyadic Green function.
Then, the two reasonings presented in this paper for an
electromagnetic and
 a scalar wave show the primacy of the Huygens Fresnel
theorem given by Eq.~(\ref{green}) for the scalar wave and by
Eq.~(\ref{supergreen}) for the electromagnetic field,
respectively.\\
A few further remarks are here relevant: First, the mathematical
results described here constitutes a justification of the physical
``trick'' introduced by Smythe and Jackson. However more work have
must be done in order to see if the method based on scalar Green
functions could be extended to other geometries. Second, the
Smythe formula allows one to express the electromagnetic field
radiated by the aperture (far-field) as a function of the
near-field existing in the aperture plane. This method could thus
be useful for calculating the field generated by a NSOM aperture
if we know the optical near-field (computed, for example, by using
numerical methods discussed in
Refs.~~\cite{Girard1,Girard2,Girard3,Bozhe}).

 \section{Conclusion}

 In this paper, we have justified the vectorial formula of Smythe
  expressing the diffracted field produced by an opening created
  in a perfectly metallic screen. Our justification is based only
  on the Huygens principle for electromagnetic wave and on the
  specifical nature of boundary conditions for the Maxwell field.
  This proof differs from the ones presented in the literature
  because it does not use the concept of current sheets introduced by
  Smythe and Jackson. The demonstration uses only the scalar Green
  function in free space and does not consider Dirichlet or Neumann
  boundary conditions as involved in the Green dyadic method.

 \appendix

 \section{}
 Let $\Psi\left(\vec{r}\right)$ be a scalar wave solution of the
 Helmoltz equation for the problem of diffraction by an opening $\delta S$ in a plane
 screen $S$. In order to define completely the problem, we must
 impose boundary conditions on the screen surface. Here, we choose $\Psi\left(\vec{r}\right)|_{S-\delta S}=0$
  for any point on the screen (Dirichlet problem). The Neumann
problem can be treated
 in a similar way. For such a problem, we can in principle always divide
 the field into an incident one, called
 $\Psi_{\textrm{inc}}\left(\vec{r}\right)$ and existing independently
 of any screen, and into a scattered field $\Psi'\left(\vec{r}\right)$, produced by sources in the
 screen. The problem cannot be solved without postulating some
 properties of the sources. A way to do this is to introduce a
 source term $J\left(\vec{r}\right)$ in the second member of
 the Helmoltz equation such that this term goes to zero rapidly
 outside of the pancake volume occupied by the screen. Then, we have
 $[\nabla^{2}+k^{2}]
 \Psi\left(\vec{r}\right)=-J\left(\vec{r}\right)$. Imposing Sommerfeld's radiation
 condition at infinity  gives us the solution
 \begin{equation}
 \Psi'\left(\vec{r}\right)=\int_{\textrm{pancake}}J\left(\vec{r'}\right)G\left(\vec{r},\vec{r'}\right)
 d^{3}{r'}.
 \end{equation}
  We deduce the important fact that this
 potentiel $\Psi'\left(\vec{r}\right)$ must be an even
 function of $z$. This is consistent with the Kirchhoff
 formula applied on the surface of Fig.~1(B). Imposing the condition $\Psi'\left(x,y,z\right)=\Psi'\left(x,y,-z\right)$
 implies
\begin{equation}
\Psi\left(\vec{r}\right) =-\int_{\left(S-\delta S\right)}
G\left(\vec{r},\vec{r'}\right){\hat{z}}\cdot\overrightarrow{\nabla'}\Psi'\left(\vec{r'}\right)\textrm{d}S'
\end{equation}
 which defines the source term $J_{S}\left(x,y\right)$ (surface
density) by $J_{S}\left(x,y\right)=-\lim_{z\rightarrow
0^{+}}{\hat{z}}\cdot\overrightarrow{\nabla}\Psi'\left(x,y,z\right)$.
It is worth noting that the even character of $\Psi'$ and the
field continuity in the aperture impose
${\hat{z}}\cdot\overrightarrow{\nabla}\Psi'\left(x,y,z=0\right)$
in the opening. In order to complete the problem, we must define
the reflected field $\Psi^{r}\left(\vec{r}\right)$ produced by the
sources when the plane screen contains no aperture. Since for
$z>0$ there is no field, we must choose
$\Psi^{r}\left(x,y,z\right)=-\Psi^{i}\left(x,y,z\right)$  in this
half plane. The requirement that the source field is  an even
function of $z$ imposes
$\Psi^{r}\left(x,y,z\right)=-\Psi^{i}\left(x,y,-z\right)$ for
$z<0$. In this form, the problem is similar to the one described
by Bouwkamp~\cite{Bouwkamp2} and it can be solved. The rest of the
reasoning is similar to the one given for the Smythe formula.
Identifying the Kirchhoff integral on the two different surfaces
represented in Figs.~2(A) and 2(B), we obtain
 \begin{widetext}
\begin{eqnarray}
\Psi'\left(\vec{r}\right) =2\int_{S^{+}}
\Psi'\left(\vec{r'}\right){\hat{z}}\cdot\overrightarrow{\nabla'}G\left(\vec{r},\vec{r'}\right)
\textrm{d}S'
+\left(\int_{S_{\infty}^{+}}-\int_{S_{\infty}^{-}}\right)[
\Psi'\left(\vec{r'}\right)\vec{n'}\cdot\overrightarrow{\nabla'}G\left(\vec{r},\vec{r'}\right)
-G\left(\vec{r},\vec{r'}\right)\vec{n'}\cdot\overrightarrow{\nabla'}\Psi'\left(\vec{r'}\right)]\textrm{d}S'.\label{rayleigh2}
\end{eqnarray}
\end{widetext}
As for the Smythe formula, we can use the symmetry properties of
the field as well as its asymptotic behavior at infinity to
transform Eq.~(\ref{rayleigh2}) into
\begin{eqnarray}
\Psi\left(\vec{r}\right) =2\int_{\delta S^{+}}
\Psi\left(\vec{r'}\right){\hat{z}}\cdot\overrightarrow{\nabla'}G\left(\vec{r},\vec{r'}\right)
\textrm{d}S'
\end{eqnarray} which is equivalent to the Rayleigh-Sommerfeld result given by
Eq.~(\ref{rayleigh}).

\newpage

Fig.~1. The problem of diffraction in electromagnetism. The
incoming wave comes from the $z<0$ half-space and is diffracted by
the aperture $\delta S$ located in the plane screen $S$ at $z=0$.
The unit vector $\vec{n}'={\hat{z}}$ used in the text is
represented.
\\

Fig.~2. The two surfaces of integration for the application of the
vectorial
 kirchhoff theorem.

\begin{figure}[h]
\includegraphics[width=3in]{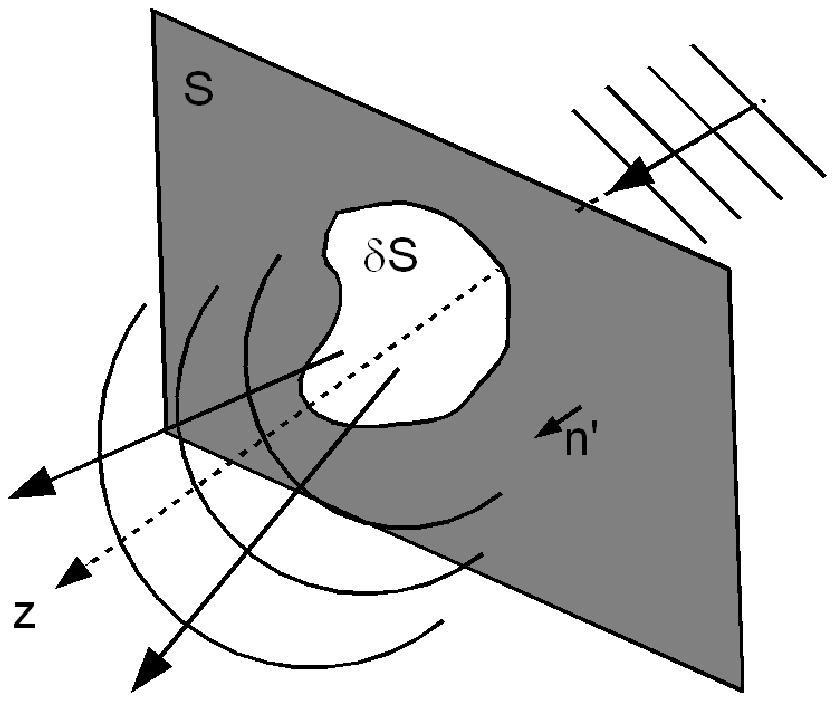}
\caption{}
\end{figure}

\begin{figure}[h]
\includegraphics[width=3in]{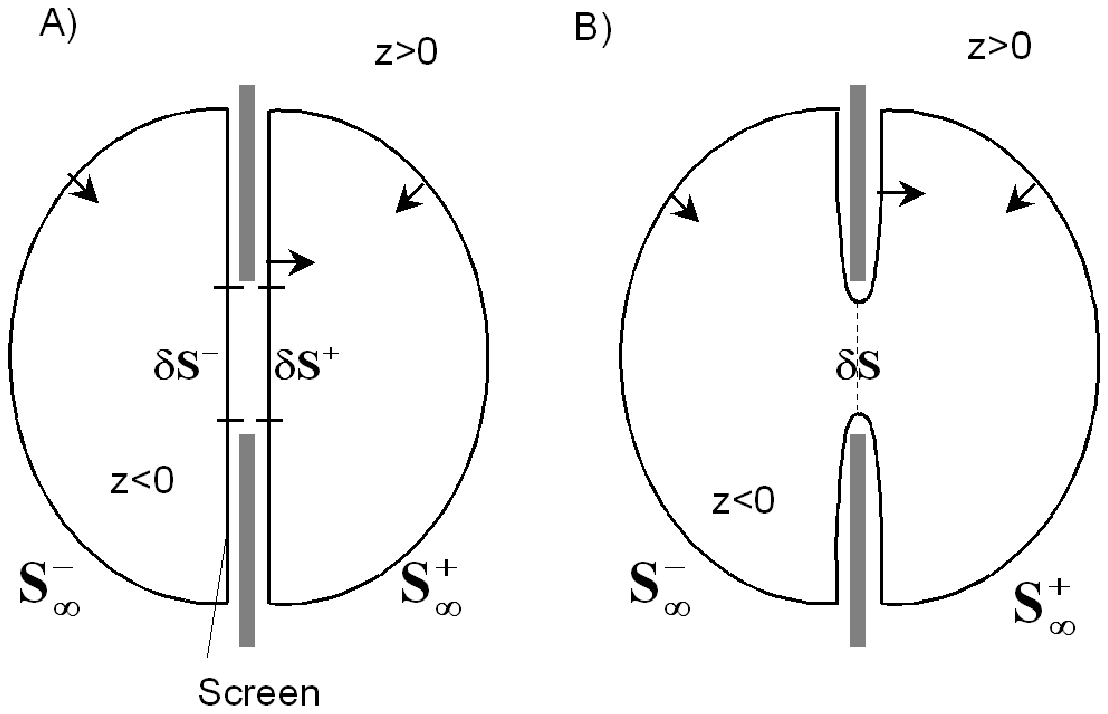}
\caption{}
\end{figure}
\end{document}